\documentclass{PoS}

\title{Consistency of neutrino DIS and the present parton distribution functions.}

\ShortTitle{Consistency of neutrino DIS and the present parton distribution functions.}

\author{\speaker{Hannu Paukkunen} and Carlos A. Salgado \\
University of Santiago de Compostela \\
Departamento de F\'\i sica de Part\'\i culas and IGFAE, Spain \\
Email: \email{hannu.paukkunen@usc.es}, \email{carlos.salgado@usc.es}}

\abstract{We study the nuclear effects in the neutrino$\backslash$anti-neutrino-nucleon
deep inelastic scattering (DIS) by comparing the NuTeV, CDHSW, and CHORUS
cross-sections from Iron and Lead targets to the predictions derived
from the latest parton distribution functions (PDFs). The nuclear modifications
found seem to display agreement with those in charged lepton DIS. Our study
thus lends support to the consistency of employing neutrino data in global
fits of PDFs.}

\FullConference{35th International Conference of High Energy Physics\\
                 July 22-28, 2010\\
                 Paris, France}

\begin{document}

\section{Introduction}

Neutrino beam induced DIS provide information about the
flavor decomposition of the PDFs that is difficult to obtain from
the other available data. The weakness of the neutrino interactions, however,
requires the usage of heavy targets --- like huge blocks of Iron or
Lead --- in the experimental setup. It is well-known by now
that the cross-sections involving bound nucleons are different
from those involving free nucleons. Founded by the series of analyses
in the past 12 years (see e.g. \cite{Eskola:2009uj,Hirai:2007sx,deFlorian:2003qf}),
such differences have turned out to be factorizable --- explainable by nuclear modifications
solely in PDFs.

Interestingly, it was reported \cite{Schienbein:2007fs} that the NuTeV $\nu {\rm Fe}$-data
\cite{Tzanov:2005kr} implies that the nuclear effects in PDFs seem to be different in the neutrino-DIS than they are
in the charged lepton DIS. Such a process-dependent difference would evidently ruin the factorization
in the case of nuclear targets, and without factorization these data would be useless
in improving the flavor decomposition of the free proton PDFs.


In this talk we summarize our analysis published in \cite{Paukkunen:2010hb} which
supports the factorization in neutrino-nucleon DIS and implies
that the result of \cite{Schienbein:2007fs} is caused by looking solely the NuTeV
data which seems to suffer from anomalous, neutrino-energy dependent normalization problems.

\section{Experimental Input}

The experimental data we use consists of neutrino$\backslash$antineutrino DIS with nuclear targets
from three independent experiments: NuTeV (Fe) \cite{Tzanov:2005kr}, CDHSW (Fe) \cite{Berge:1989hr},
and CHORUS (Pb) \cite{Onengut:2005kv}. Importantly, we employ the published cross-sections instead of the
structure functions extracted by the collaborations. These data sets provide enough kinematical overlap
to also explore the mutual consistency of the different sets.

\section{Theoretical Framework}

Our analysis employs the $\overline{MS}$ variable flavor number scheme, its SACOT-prescription.
The baseline free proton PDFs are taken from CTEQ6.6 \cite{Nadolsky:2008zw}, and the nuclear
modifications from \cite{Eskola:2009uj}. In addition, we account for the target mass (TM) correction,
and correction for electroweak radiation (RAD) when calculating the cross-sections (see \cite{Paukkunen:2010hb} 
for details).
This is, in fact, the main reason why we prefer the cross-section data instead of the extracted
structure functions: Both of these corrections clearly depend on the employed set of PDFs
(see Table~\ref{Table:chi2values}), and consequently the structure functions provided by the experiments depend on what
they assumed about them.

\section{$\chi^2$-Values}

\begin{table}
\begin{center}
{\footnotesize
\begin{tabular}{ccc}
 Radiative and Target Mass corrections & CTEQ6.6 &  CTEQ6.6$\times$EPS09 \\
\hline
\hline
 NuTeV	& 1.51  & 1.05 \\
 CHORUS & 1.15  & 0.79 \\
 CDHSW  & 1.10  & 0.71 \\
 \\
No Radiative or Target Mass corrections & CTEQ6.6 &  CTEQ6.6$\times$EPS09 \\
\hline
\hline
 NuTeV	& 1.35  & 1.08 \\
 CHORUS & 1.23  & 1.09 \\
 CDHSW  & 0.96  & 0.86  \\
\end{tabular}
}
\caption[]{\small The $\chi^2/N$-values computed using CTEQ6.6 with and without nuclear
modification from EPS09. The numbers are given for calculations with and without the radiative
and the target mass corrections.}
\label{Table:chi2values}
\end{center}
\end{table}

In Table~\ref{Table:chi2values} we provide the $\chi^2/N$-values measuring the
agreement between the data and the theory.
Evidently, in the case of CDHSW and CHORUS data, the full calculation with the
CTEQ6.6 and the EPS09 gives excellent results with $\chi^2/N < 0.8$. Also, the addition 
of the TM and RAD corrections improve the results. Only the NuTeV
data behaves differently: Apart from good $\chi^2/N$, it appears
strange why there is practically no effect whether we apply RAD and TM corrections or not. 

\section{Shape Of The Nuclear Modifications}

Only from the bare $\chi^2/N$-values displayed in Table~\ref{Table:chi2values}, a statistician
would say that the calculations are in agreement with the data even without the nuclear corrections
in PDFs. 
While probably true, the shape of the data --- plotted as a function of Bjorken-$x$ --- clearly shows the
existence of the nuclear modifications. We plot two different ratios of cross-sections:
\begin{equation}
 R^{\rm CTEQ6.6} \equiv \frac{\sigma^{\nu,\overline\nu}\left({\rm Experimental} \right)}{\sigma^{\nu,\overline\nu}\left({\rm CTEQ6.6} \right)}, \qquad R^{\rm CTEQ6.6 \times EPS09} \equiv \frac{\sigma^{\nu,\overline\nu}\left({\rm CTEQ6.6 \times EPS09} \right)}{\sigma^{\nu,\overline\nu}\left({\rm CTEQ6.6} \right)} \label{eq:nuclear_mod_npdfs}.
\end{equation}
In $R^{\rm CTEQ6.6}$ the denominator is stripped from the nuclear effects, and this ratio should
therefore reflect the nuclear effects present in the experimental data. The other ratio
$R^{\rm CTEQ6.6 \times EPS09}$ is purely theoretical one and should agree with the former one
if the nuclear effects in PDFs are universal.

\begin{figure}[!htb]
\center
\hspace{-1.2cm}
\includegraphics[scale=0.45]{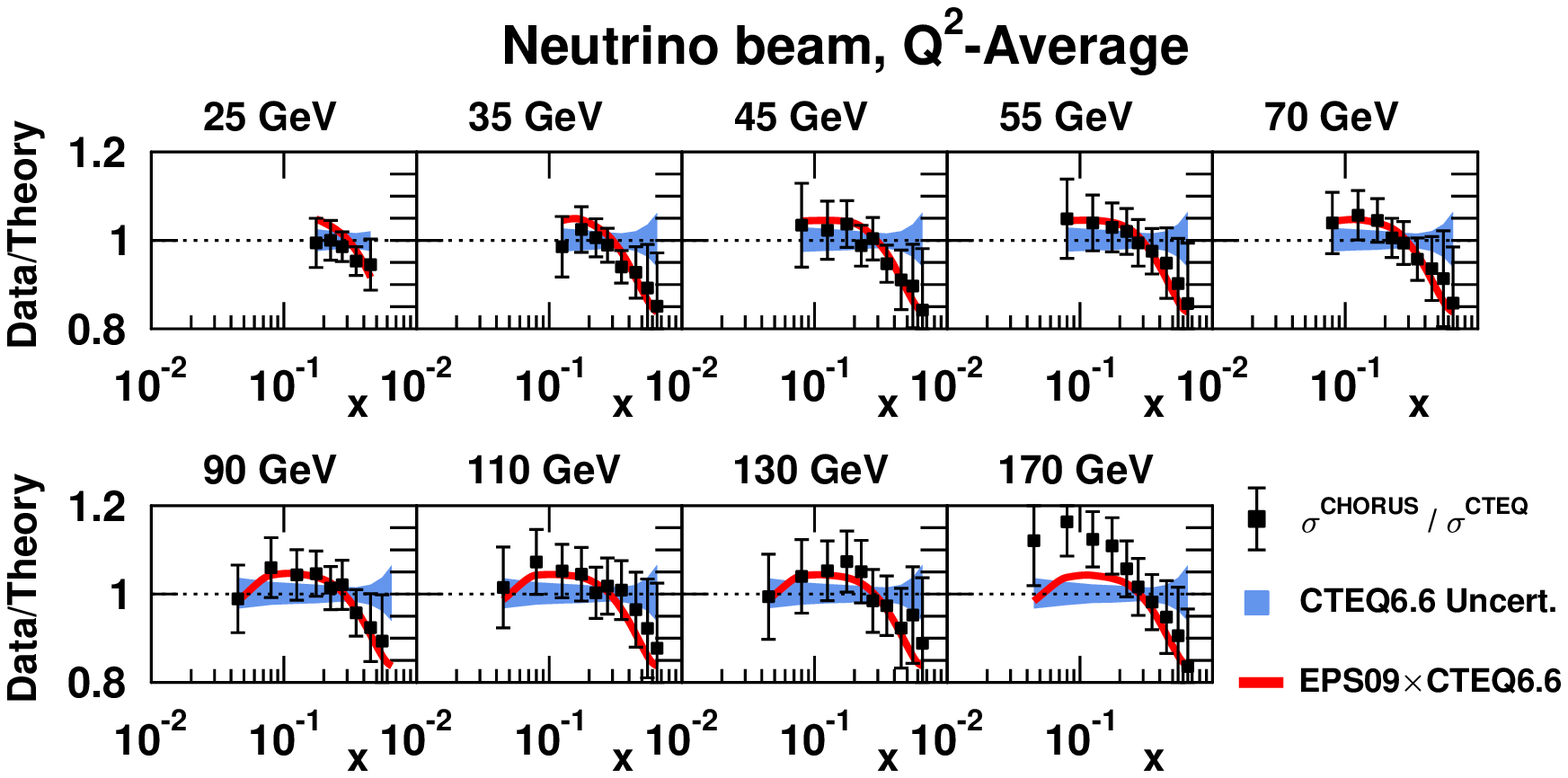} 
\hspace{-0.2cm}
\includegraphics[scale=0.45]{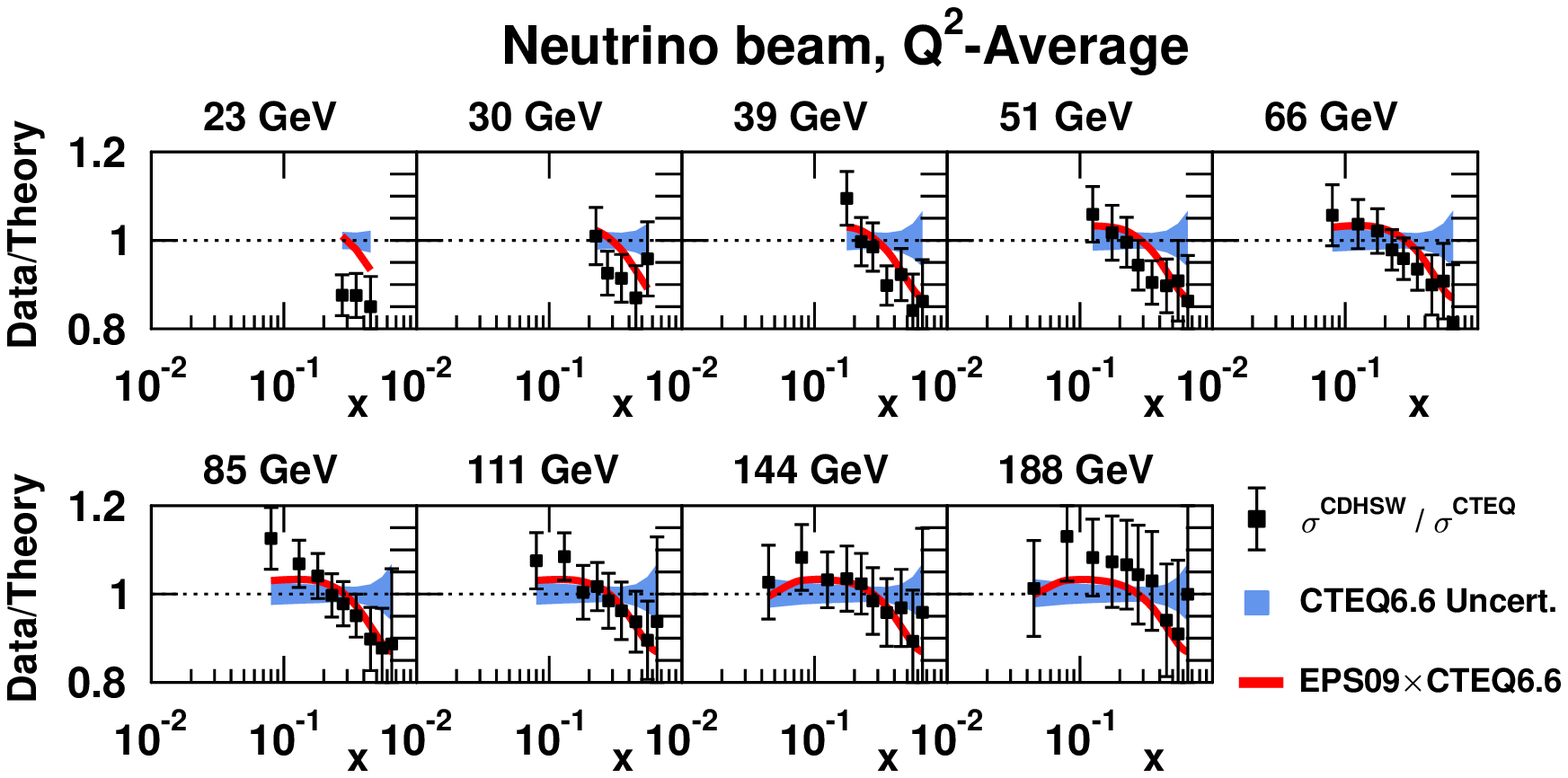} 
\includegraphics[scale=0.47]{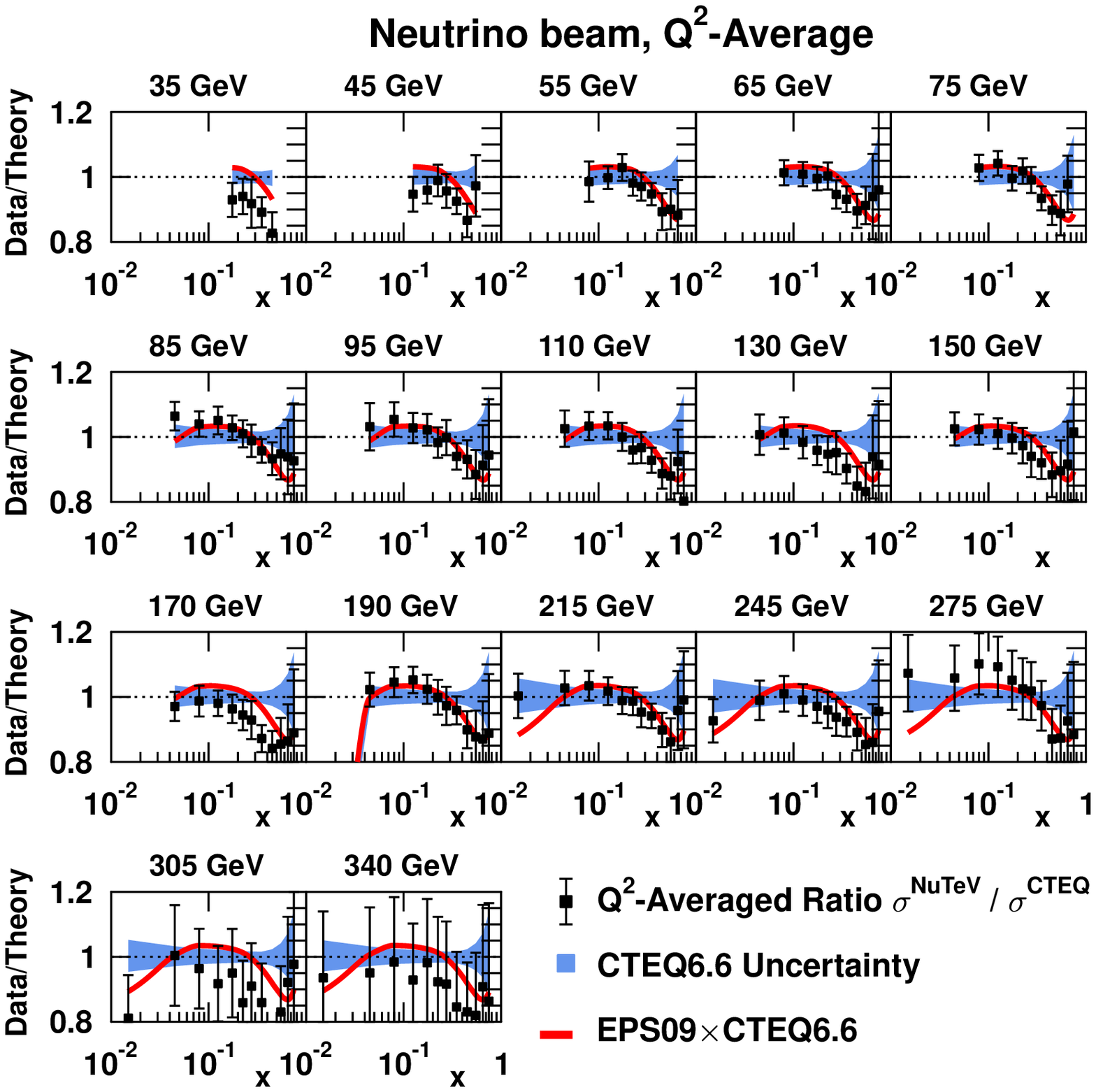}
\caption[]{\small The $Q^2$-averaged CHORUS, CDHSW and NuTeV neutrino data.}
\label{Fig:Q_nu_average_NuTeV}
\end{figure}

In Figure~\ref{Fig:Q_nu_average_NuTeV}, we plot the $Q^2$-averaged versions of these
ratios as a function of $x$ for different neutrino energies $E_{\rm beam}$. We show
the case of neutrino beam for all three data sets omitting the figures for antineutrinos 
as they are similar but with substantially larger uncertainties. The nuclear effects are clearly
visible in the data: There is an excess around $x \cong 0.1$, the antishadowing,
followed by a suppression at larger $x$, the EMC-effect. This shape is generally well reproduced
by the nuclear effects from EPS09. However, in the NuTeV data there are evident, neutrino energy
dependent fluctuations in the data. For example, the normalization of the data in panels with $E_{\rm beam} = 130 \,{\rm GeV}, 170 \,{\rm GeV}, 245 \,{\rm GeV}$ evidently deviate from the predictions, while e.g data in panels
with $E_{\rm beam} = 85 \,{\rm GeV}, 95 \,{\rm GeV}, 190 \,{\rm GeV}$ are in a perfect
agreement.

As the theoretical predictions depend only very weakly on the incident neutrino energy,
these differences are impossible to fix only by changing the PDFs. That is, if the NuTeV data is
used to extract the nuclear effects, the result will be a some kind of compromise displaying tension
between different subsets of data. This should be kept in mind when 
interpreting the results.

\vspace{-0.3cm}
\section{Summary}

As a summary, we argue that the nuclear effects in PDFs extracted from other
processes than neutrino DIS, agree also with the present neutrino data - only
the NuTeV data, containing internal inconsistecies, display a tendency to different
effects. Futurte neutrino data as e.g. those from the NOMAD collaboration
may eventually help in settling down the issue.


\begin{thebibliography}{99}
  
\bibitem{Eskola:2009uj}
  K.~J.~Eskola, H.~Paukkunen and C.~A.~Salgado,
  JHEP {\bf 0904} (2009) 065
  [arXiv:0902.4154 [hep-ph]].

\bibitem{Hirai:2007sx}
  M.~Hirai, S.~Kumano and T.~H.~Nagai,
  Phys.\ Rev.\  C {\bf 76} (2007) 065207
  [arXiv:0709.3038 [hep-ph]].

\bibitem{deFlorian:2003qf}
  D.~de Florian and R.~Sassot,
  Phys.\ Rev.\ D {\bf 69} (2004) 074028
  [arXiv:hep-ph/0311227].

\bibitem{Schienbein:2007fs}
  I.~Schienbein, J.~Y.~Yu, C.~Keppel, J.~G.~Morfin, F.~Olness and J.~F.~Owens,
  Phys.\ Rev.\  D {\bf 77} (2008) 054013
  [arXiv:0710.4897 [hep-ph]].

\bibitem{Tzanov:2005kr}
  M.~Tzanov {\it et al.}  [NuTeV Collaboration],
  Phys.\ Rev.\  D {\bf 74} (2006) 012008
  [arXiv:hep-ex/0509010].

\bibitem{Paukkunen:2010hb}
  H.~Paukkunen and C.~A.~Salgado,
  JHEP {\bf 1007} (2010) 032
  [arXiv:1004.3140 [hep-ph]].

\bibitem{Berge:1989hr}
  J.~P.~Berge {\it et al.},
  Z.\ Phys.\  C {\bf 49} (1991) 187.

\bibitem{Onengut:2005kv}
  G.~Onengut {\it et al.}  [CHORUS Collaboration],
  Phys.\ Lett.\  B {\bf 632} (2006) 65.

\bibitem{Nadolsky:2008zw}
  P.~M.~Nadolsky {\it et al.},
  Phys.\ Rev.\  D {\bf 78} (2008) 013004
  [arXiv:0802.0007 [hep-ph]].


\end{thebibliography}
\end{document}